# Tunable bilayer dielectric metasurface via stacking magnetic mirrors

Hao Song, Binbin Hong*, Yanbing Qiu, Kuai Yu, Jihong Pei and Guo Ping Wang*

*College of Electronics and Information Engineering, Shenzhen University, Shenzhen, 518060, P. R. China*
*Email: b.hong@szu.edu.cn, gpwang@szu.edu.cn

**Abstract**

Functional tunability, environmental adaptability, and easy fabrication are highly desired properties in metasurfaces. Here we provide a tunable bilayer metasurface composed of two stacked identical dielectric magnetic mirrors, which are excited by the dominant electric dipole and other magnetic multipoles, exhibiting nonlocal electric field enhancement near the interface and high reflection. Differ from the tunability through the direct superposition of two structures with different functionalities, we achieve the reversible conversion between high reflection and high transmission by manipulating the interlayer coupling near the interface between the two magnetic mirrors. The magnetic mirror effect boosts the interlayer coupling when the interlayer spacing is small. Decreasing the interlayer spacing of the bilayer metasurface leads to stronger interlayer coupling and scattering suppression of the meta-atom, which results in high transmission. On the contrary, increasing the spacing leads to weaker interlayer coupling and scattering enhancement, which results in high reflection. The high transmission of the bilayer metasurface has good robustness due to that the meta-atom with interlayer coupling can maintain the scattering suppression against adjacent meta-atom movement and disordered position perturbation. This work provides a straightforward method (i.e. stacking magnetic mirrors) to design tunable metasurface, shed new light on high-performance optical switches applied in communication and sensing.

## 1. Introduction

Metasurfaces usually consist of sub-wavelength meta-atoms arranged on a two-dimensional (2D) plane for electromagnetic (EM) wavefront manipulation [1,2], which have been applied in integrated optics, information communication, on-chip photonics, and also obtained many exotic physical phenomena [3-5], new laws [6], and high-performance functional devices [7-10].

High refractive-index dielectric structures with lossless and high efficiency are important branches of metasurface research, which realize full wavefront manipulation by controlling the Mie resonances of meta-atoms via changing refractive index, geometric shape/size, et al [11-15]. Moreover, strong artificial magnetic responses can be easily obtained by simple dielectric particles [2,13,16,17]. A recent remarkable achievement based on the induced significant magnetic responses is magnetic mirrors (MMs) [18-21]. MMs exhibit in-phase high reflection near the medium interfaces, effectively eliminating the half-wave loss of electric field existing at the interfaces of the conventional metal mirrors, resulting in the enhanced large-scale near-field electric field and light-matter interactions, applied in subwavelength imaging [22], molecular fluorescence [23], Raman spectroscopy [24], perfect reflectors [25,26], etc. Up to now, it has been found that the electric responses and multiple response interferences can also generate the MMs [27-29]. Yet there are few studies on the features and applications of stacking MMs.

To address the challenges of real-time, tunable, multifunctional in emerging applications, such as computational imaging, ultrafast light, quantum communication, opto-thermal management, dynamic sensing, it provides an opportunity to develop the next generation tunable metasurfaces [30-32]. One way is to apply extra operations generally including mechanical motion [33], electric voltage [34], electrochemistry [35,36], encoding [37], liquid crystals [38], and phase change materials [39]. On the other hand, stacking a few single-layer metasurfaces is also an effective and controllable way to design tunable metasurfaces, which usually consist of two or three layers [40-42] exhibiting novel properties and applications, e.g., reflection/transmission and absorption switchers [39,43], full-space controls of EM waves [44,45], wideband chromatic aberration-free meta-mirrors [46], intelligent rapid holographic imager [47], beam splitting [48], etc. However, most multifunctional stacked metasurfaces are based on the superposition of carefully patterned single-layer metasurfaces with different functions, which lead to the harder fabrication of complex structures requiring precise alignment, and probably introduce more EM loss [32,42,49,50] and instability.

Here we present a tunable bilayer metasurface simply stacked by two identical single-layer MMs composed of air-$Al_2O_3$ dielectric concentric cylinders. The MM is excited by the interference of the dominant electric dipole resonance and other magnetic multipoles. The dynamical conversion between high reflection and transmission is realized easily by adjusting the interlayer spacing. Different from the tunability originating from the in-plane coupling between adjacent meta-atoms [51,52], the physical root of this tunable metasurface is the controllable interlayer coupling leading to controllable scattering of the composite meta-atoms. Note that interlayer coupling corresponds to scattering suppression and high transmission, whereas decoupling corresponds to scattering enhancement and high reflection. The near-field electric field enhancement of MM plays a large role in interlayer coupling when two layers are close. High transmission of the bilayer metasurface is robust because the scattering suppression has great tolerance to the disturbance of positional disorders and adjacent meta-atom movements. Our finding is verified in microwave experiments. This work may open a new pathway to the research and application of stacking magnetic mirrors and tunable metasurfaces, and probably provide a promising component for high-performance optical switching applied in communication and sensing.

## 2. EM responses of metasurfaces
### 2.1 Single-layer magnetic mirror

We begin by considering a single-layer metasurface composed of infinite concentric cylinders periodic arranged along the *x*-direction with the minimal face spacing *p* depicted in Fig. 1(a). The radius of the air-core is $r_0$, the shell is $Al_2O_3$ with radius $r_1$ and relative permittivity $\varepsilon_1 = 9.424 + i2.920 \times 10^{-3}$ [53]. The incident EM wave vector **k** (wavenumber *k*) in the air is perpendicular to the centreline of the cylinder and the polarization of electric field **E** is along the *z*-axis. We set the fixed incident frequency of 10 GHz (incident wavelength λ) and parameter $kr_1 = 1.593$ (i.e. $r_1$ fixed). The $\alpha = r_0 / r_1$ is to characterize the inner radius. Throughout this paper, the numerical simulations are completed by the commercial software COMSOL [54]. The whole structure is immersed in free space and analyzed using a 2D model with period and perfectly matched layer (PML) boundaries.

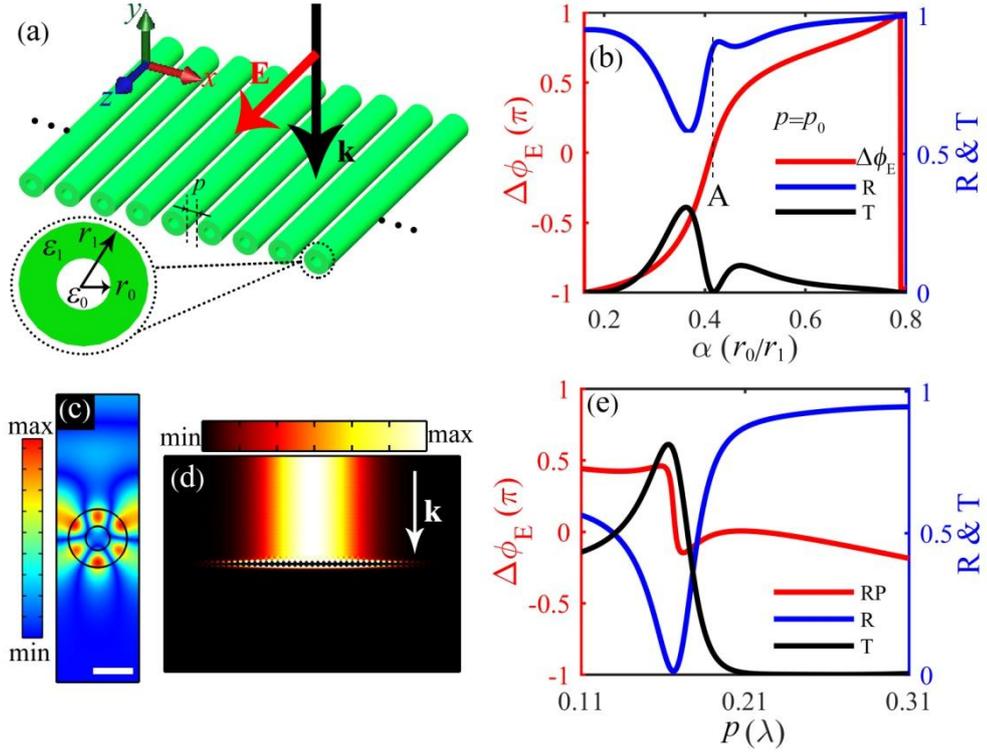

Figure 1. (a) Schematic of a single-layer periodic metasurface and a normally incident linearly polarized plane wave. (b) Reflectivity (R, blue), transmissivity (T, black), reflection phase $\Delta\phi_E$ curves of the single-layer metasurface with $p=p_0$. Point A corresponds to the magnetic mirror with $\Delta\phi_E=0$. (c) Normalized electric field $|\mathbf{E}|$ distribution at point A. Hereafter, the scale bar denotes 10 mm. (d) Energy distribution at point A under Gaussian beam incidence. (e) Spectra of R (blue), T (black), $\Delta\phi_E$ (red) curves of the metasurface at point A with respect to $p$.

With $p=p_0$ (0.21$\lambda$) and $\alpha$ varying, the reflectivity (R), transmissivity (T), reflection phase ($\Delta\phi_E$) of the metasurface are demonstrated in Fig. 1(b). $\Delta\phi_E$ is the phase difference between the reflected and incident electric field at the interface plane, which is tangent to the top of the metasurface where the incident wave directly interacts with. In the discussed range of $\alpha$, $\Delta\phi_E$ presents the continuous change of full phase accompanied by the variable R and T. Interestingly, a MM with $\Delta\phi_E=0$, R=0.87, and T=0.01 is achieved at point A (corresponds to $\alpha$=0.414).

Figure 1(c) shows the normalized electric field $|\mathbf{E}|$ distribution of one unit cell of the MM. The interface plane is located near the maximum value of stand wave interference of the electric field. $|\mathbf{E}|$ is mainly confined in the shell and the pattern in the cylinder indicates the presence of magnetic octupole (MO) [55]. Figure 1(d) displays the energy density distribution of the MM under a Gaussian beam with wave vector $\mathbf{k}$. One can see that a little partial energy is localized near the interfaces and most are reflected.

The period dependence of the EM response of the MM is shown in Fig. 1(e). The metasurface corresponding to point A maintains the MM effect ($\Delta\phi_E\sim0$, R>0.8) in the range of $p$ from 0.20$\lambda$ to 0.23$\lambda$, displaying a perfect conversion of R without T to T without R in the whole range. For simplicity, we define the perfect conversion efficiency ($\eta$) of the R without T to T without R as the absolute value of the maximum value ratio of T to R, i.e. $\eta = |T_m / R_m|$. According to the $\eta$, we can estimate whether there is EM energy dissipation. Here, the $\eta$ is about 0.89, which indicates the reflected energy converts into the transmission and small partial EM dissipation. Moreover, the MM has an asymmetric transmissivity curve near the incident frequency [see Fig. 3(a)].

**2.2 Tunable bilayer metasurface**

Figure 2(a) depicts a schematic of a bilayer metasurface stacked by two identical MMs corresponding to point A. The two MMs are aligned and $h$ is the minimum interlayer face spacing. **k** is perpendicular to the metasurface and **E** is parallel to the centerline of the cylinder. Adjusting $h$ continuously, Fig. 2(b) shows dynamic R (red) and T (blue) curves of the bilayer metasurface. Point B corresponds to the high T without R and point C denotes any point of the high R without T in the whole range. Here, $h$ of point B equals $h_0$ (0.302λ). The $\eta$ is about 1.02, which indicates the bilayer metasurface can realize nearly equal intensity perfect interconversion between high reflection and high transmission. Furthermore, the bilayer metasurface obtains high transmission with larger operating bandwidth compared with Fig. 1(e). The asymmetric peak near $h$=0.56λ indicates the existence of 1st order Fabry-Perot (FP) resonance between the two layers, the other asymmetric peaks at larger $h$ show the presence of high-order FP resonances.

The insets in Fig. 2(b) show the |**E**| distributions of points B and C with incident wavevector **k**, respectively. Here, we randomly selected a point C with $h$=1.20λ. The two cylinders occur obvious coupling at point B while decoupling at point C. Figures 2(c) and 2(d) show the energy distributions of the bilayer metasurface under a normal incident Gauss beam of points B and C, respectively. As a result, point B corresponds to the turn 'on' state of incident energy almost without loss as shown in Fig. 2(c), whereas the turn 'off' state with total reflection at point C is presented in Fig. 2(d).

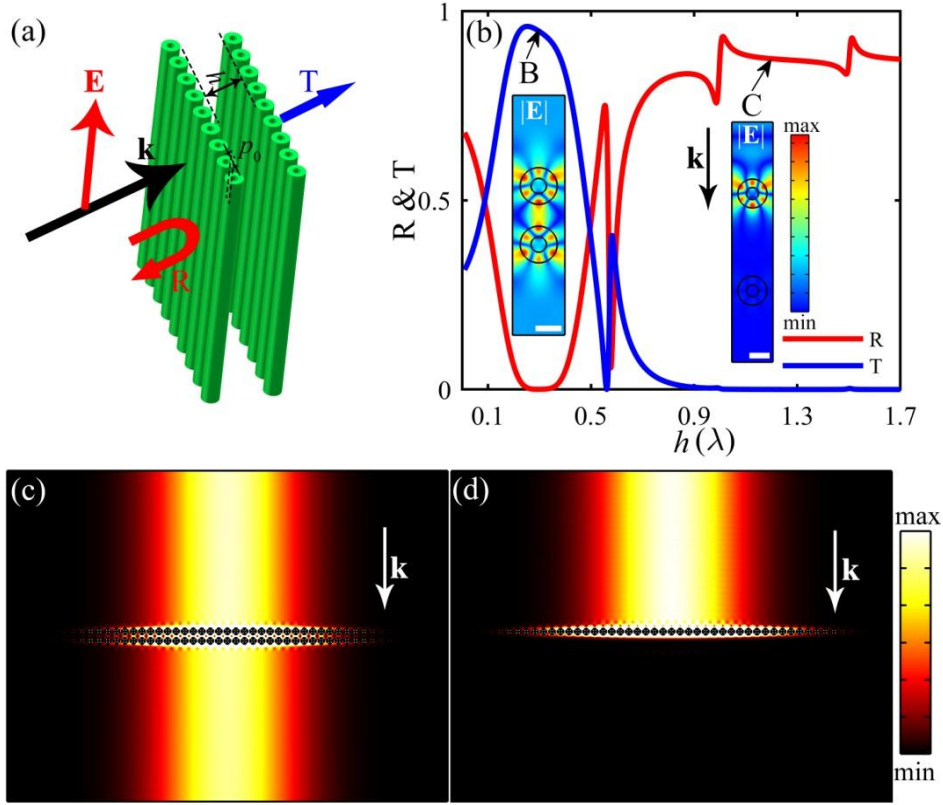

Figure 2. (a) Schematic of a bilayer metasurface stacked by the two MMs corresponding to point A with minimal interlayer face spacing $h$ under the same incident plane wave. (b) R and T spectra of the bilayer metasurface as a function of $h$. The insets correspond to the |**E**| distributions of points B and C, respectively. (c) and (d) energy density distributions of the metasurface under a Gaussian beam corresponding to points B and C, respectively.

## 3. Experimental verification

We verify this intriguing phenomenon of the tunable metasurface in the transmission geometry via microwave experiments. The schematic is shown in the inset of the top right corner in Fig.4. A microwave vector network analyzer (VNA) with an aligned transmitter and receiver generates a polarized EM wave with a continuously variable frequency at about 10GHz. The measurement has been calibrated using the thru response calibration, which treats the free transmission between the transmitting and the receiving horns without any obstacle in between as the reference signal. The periodic single-layer (PS) MM at point A is constructed by 12 finite length alumina ceramic rods with the spacing $P_0$, where the length is about $6.67\lambda$ as shown in Fig. 3(c). The bilayer metasurface is stacked by the two PS MMs. Figure 3(a) shows the frequency response of the MM at point A. It exhibits an asymmetric T curve and zero transmission near 10 GHz. Figure 3(b) is the frequency response in the experiment. According to the minimum value and asymmetric profile of the T, we confirm that the response frequency of the MM redshifts to 9.853GHz. The metasurface samples are located in the far-field domain for approximate plane wave incidence. The distances between the sample and the transmitter and receiver are based on the antenna far-field condition $d \geq 2D^2/\lambda_0$, where D is the maximum linear dimension of the antenna and $\lambda_0$ is the operating wavelength. The supporting low-density expanded polystyrene foams are almost transparent to the operating microwave, which can minimize their interference with the signal.

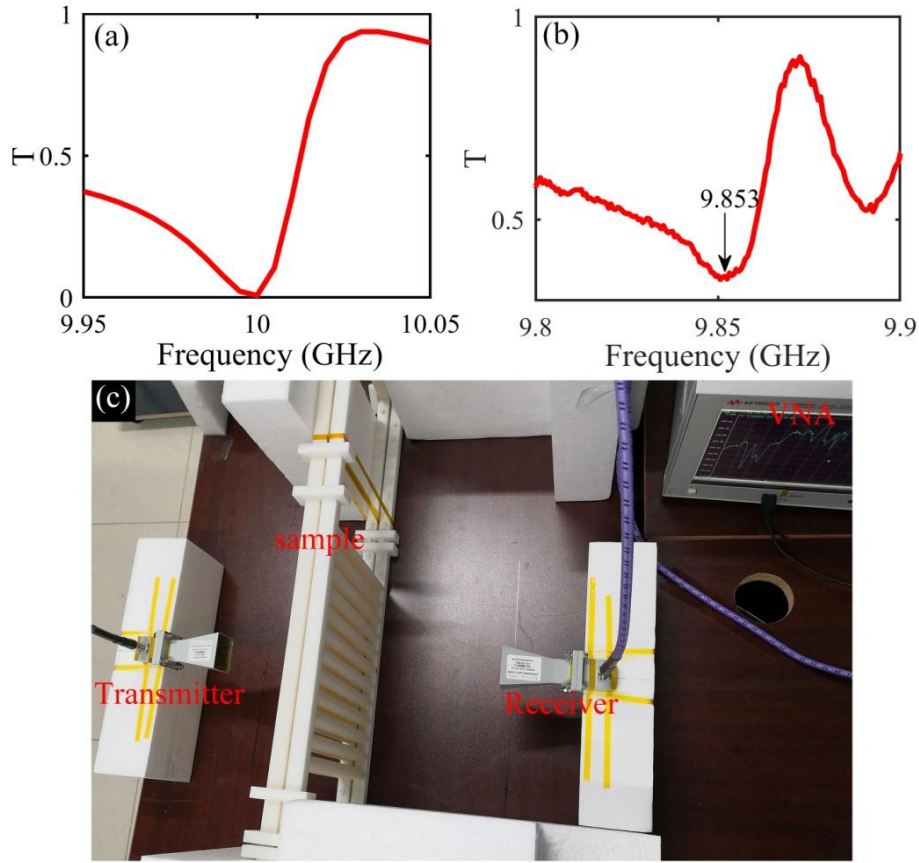

Figure 3. T spectra of frequency dependency of the single-layer MM (a) in the simulation and (b) in the experiment. (c) Experimental setup.

The experiment starts with the MM. Its effective response frequency redshifts to 9.853GHz [see

Fig. 3(b)] corresponding to the free-space wavelength of $\lambda_1$. The T of PS MM is represented by a red circle in Fig.4. Each point and the relevant error bar correspond to the average value and standard deviation of multiple measurements, respectively. Therefore, the R of the MM is about 0.64 because T is about 0.36 and the absorption of the material is negligible. The red dots in Fig. 4 denote the T spectrum of the bilayer periodic (biP) metasurface as $h$ varies, where the red dotted line is achieved by data interpolating and fitting. Thus, the biP metasurface exhibits a tunable T with a range of about from 0.07 to 0.90, which agrees well with the blue curve of numerical simulation in Fig. 2(b). The peak near $h=0.52\lambda_1$ originates from the FP resonance. However, the $h$ of the maximum T becomes smaller than $h_0$ due to effective response frequency change. The deviation between experiment and simulation comes from the limitations of experimental conditions (e.g., finite-length cylinders, finite width array) and fabrication precision. Notwithstanding these limitations, the experiment does verify the tunability of the bilayer metasurface consistent with the theoretical simulation.

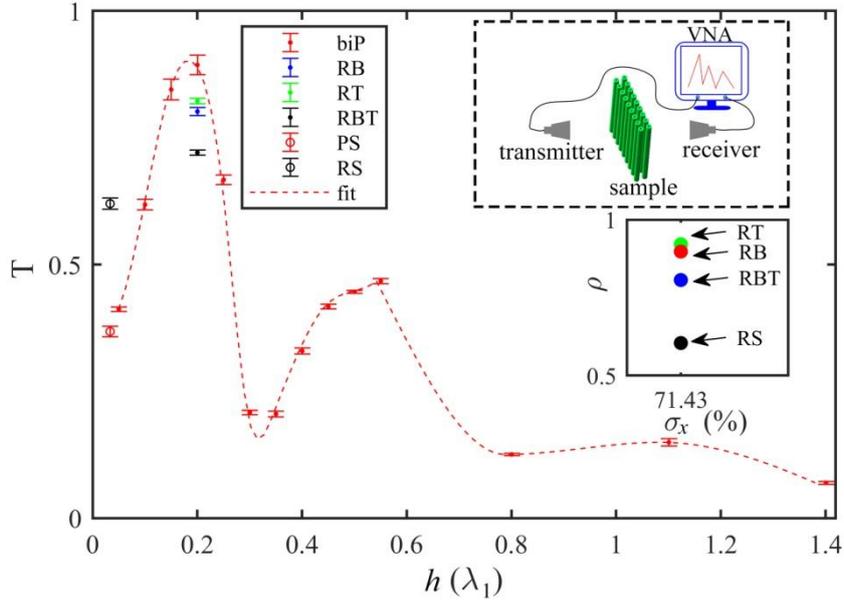

Figure 4. T spectra of metasurfaces to $h$ in microwave experiment. $\lambda_1$ corresponds to the actual effective response wavelength in the experiment. biP: bilayer periodic metasurface; RB: bilayer metasurface with the random arrangement in the bottom layer; RT: the random arrangement in the top layer; RBT: the random arrangement in the two layers; PS: periodic single-layer metasurface; RS: random single-layer metasurface. The red dotted line is the interpolating fitting of data. Each point and the related error bar correspond to the average value and standard error of multiple measurements, respectively. The inset in the top right corner shows the experiment setup. VNA refers to a microwave vector network analyzer. The other inset below presents the response factors ($\rho$) of random metasurfaces.

## 4. Scattering analysis
### 4.1 Scattering of an isolated cylinder

To further understand the underlying physical mechanisms of the tunability of the bilayer metasurface, we next consider the scattering properties of a standalone infinitely-long hollow-core cylinder under a polarized plane wave, as depicted in the inset of the top left corner in Fig. 5(a). Here, one cylinder is immersed in free space and the PML replaces outer free space in the simulation. According to the Mie theory [56], the absorption (abs), scattering (sca), extinction (ext) efficiencies as functions of $\alpha$ are shown in Fig. 5(a). The sca (blue) and ext (red) curves are completely uniform and the EM absorption can be ignored. Moreover, point A is located near the 1st peak of the sca curve. The inset of the lower right corner shows the scattering electric field |$\mathbf{E}_s$| distribution of point A.

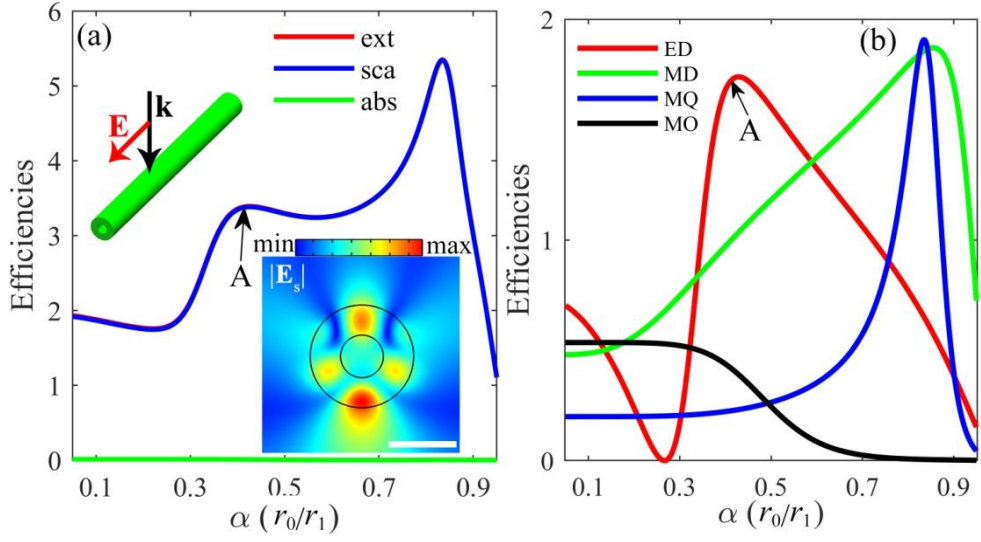

Figure 5. (a) Scattering (sca), absorption (abs), extinction (ext) efficiencies of an isolated cylinder. Geometry and polarized plane EM wave excitation are displayed in the top left corner. Normalized scattering electric fields |$\mathbf{E}_s$| at point A is shown in the lower right corner. (b) Scattering efficiencies (SEs) of spherical multipoles of the isolated cylinder.

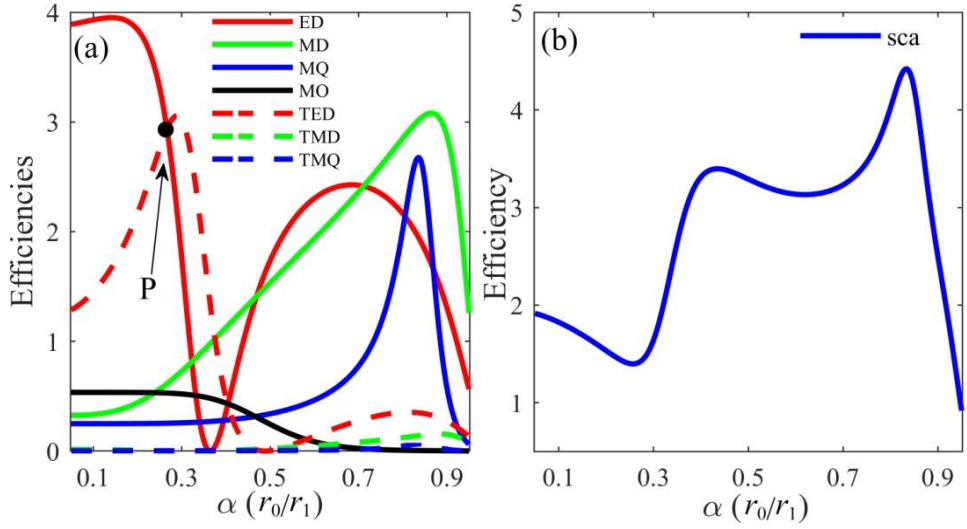

Figure 6. (a) Scattering contributions of cartesian multipoles of an isolated cylinder. (b) SE is obtained by the summation of spherical multipoles contributions.

To reveal the principle of the MM, the semianalytical scattering multipole decomposition can be carried out in Cartesian bases based on displacement current density expressed as [57]

$$\mathbf{J}(\mathbf{r}) = -i\omega\varepsilon_0[\varepsilon(\mathbf{r})-1]\mathbf{E}(\mathbf{r}), \tag{1}$$

where $\mathbf{r}$ is local coordinate, $\omega$ is the incident angular frequency, $\varepsilon_0$ is vacuum permittivity, $\varepsilon(\mathbf{r})$ is position-dependent relative permittivity, and $\mathbf{E}(\mathbf{r})$ is the total electric field. The basic Cartesian multipole moments include [58-60], the electric dipole (ED), magnetic dipole (MD), magnetic quadrupole (MQ), MO. In addition, the little contribution of higher-order electric multipoles will be ignored [55]. The corresponding toroidal moments of residual terms of the basic Cartesian multipole moments aforementioned are also considered [59-62], i. e. the toroidal electric dipole (TED), toroidal magnetic dipole (TMD), toroidal magnetic quadrupole (TMQ).

The scattering efficiency (SE) of the basic Cartesian multipoles and toroidal moments are shown in Fig. 6(a). One can see that abundant Cartesian multipoles are excited in an air-$Al_2O_3$ cylinder, especially the high-order magnetic responses MQ and MO. However, the scattering contributions of toroidal moments are little except for the TED. Interestingly, point P shows the equal contributions of Cartesian ED and TED, i.e. an anapole mode generated [63]. Furthermore, a spherical multipole moment is calculated by the interference between the basic Cartesian multipole moment and the corresponding toroidal moment [59]. Figure 5(b) shows the obvious resonant SE curves of spherical ED (red), MD (green), MQ (blue), and the nonresonant curve of spherical MO (black). Point A is very close to the ED resonance and the resonances of the MQ and MD are overlapped. Summating the contributions of all discussed spherical multipoles can achieve an effective approximate curve [see Fig. 6(b)] of the total SE curve (blue, in Fig. 5(a)). Therefore, it can confirm the existence of MO mode at point A consistent with the |$E_S$| distribution. The interference of dominant ED resonance and other magnetic multipoles have generated the MM corresponding to point A.

**4.2 Scattering of multiple cylinders**

Continually, we investigate the scattering properties of multiple cylinders. Figure 7(a) shows the SEs of N (N=2,⋯,7) identical cylinders arranged along the *x*-direction with equal spacing $p_0$ and the same incident excitation aforementioned. As N increases, the SEs at point A change slightly because of the weak coupling between cylinders, the peak locations scarcely moving is due to the strong ED resonance of each cylinder. Consequently, the SEs at point A are hardly affected by multiple scattering.

Figure 7(b) shows the SE of two identical cylinders arranged along the *y*-direction with spacing $h_0$. As a result, the SE of point A is significantly less than the cases of a single cylinder (blue curve in Fig. 5(a)) and two arranged in the *x*-direction (green curve in Fig. 7(a)). The inset shows the |$E_s$| distribution of point A. One can see obvious coupling between the two cylinders, which is the reason for the SE decrease. Furthermore, a smaller scattering cross-section means less interaction between incoming wave and particle resulting in more transmission.

For discussing device response, it is necessary to study the scattering influence of relative position change between adjacent units. Each composite unit is constituted by two identical cylinders with spacing $h_0$ and moving synchronously see the red dashed frame in Figs. 7(c) and 7(d). Let us center attention on the cylinders at point A. Figure 7(c) shows SEs of identical two cylinders (N2, blue) and two composite units (N4, red) when the spacing *p* changes in the *x* direction. The N2 and N4 have almost the same SE when *p*∼0, while N4 decreases significantly at point D with *p*=$p_0$ and the gentle range is much wider near this point. Therefore, near point D, the composite unit can much better suppress the scattering and maintain this feature even if there is a relatively horizontal position shift of the adjacent unit.

Figure 7(d) shows the SEs of N2 and N4, adjacent units having spacing $p_0$ in the *x*-direction and moving oppositely in the *y*-direction, where the composite unit has fixed spacing $h_0$. The $\Delta y$ is the center spacing in the *y*-direction of two cylinders or composite units, which have the same initial heights. In the entire range of $\Delta y$, two curves exhibit a similar trend yet N4 has a much lower SE. Consequently, the composite unit can also much greater suppress the scattering and maintain this feature although with the relative movement in the *y*-direction.

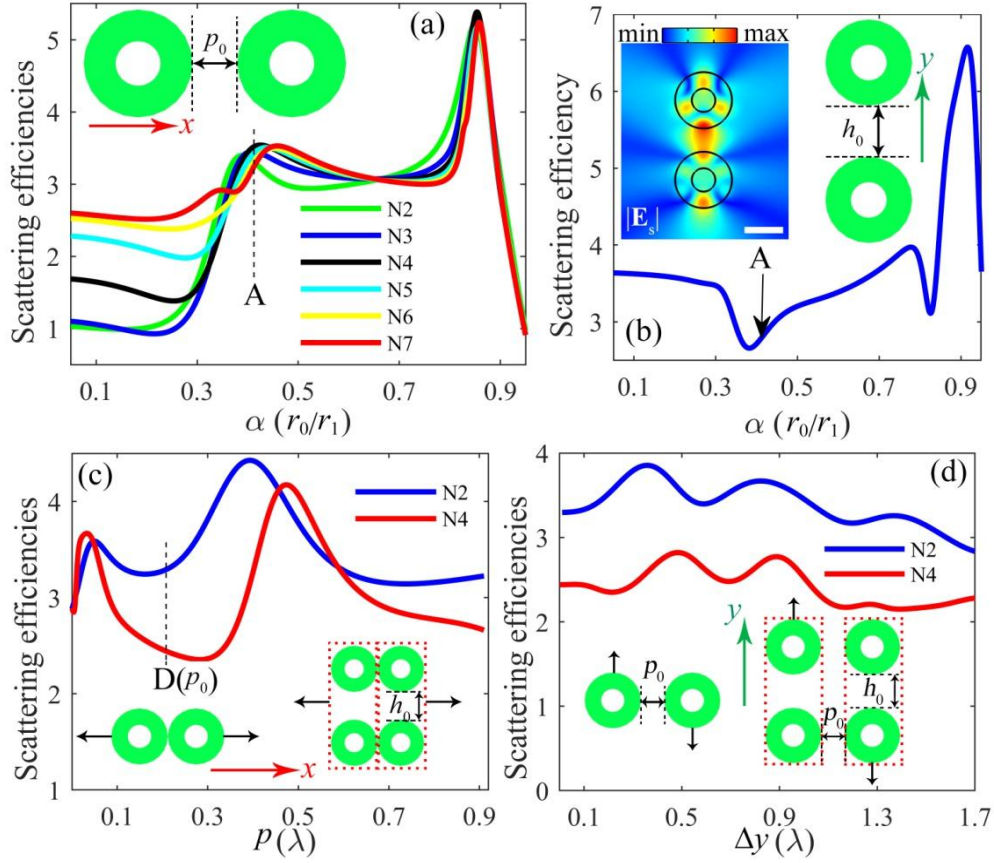

Figure 7. (a) SEs of N identical cylinders arranged in the x-direction with the same $p_0$. (b) SE of two identical cylinders arranged in the y-direction with the same $h_0$. $|E_s|$ at point A is shown in the top right corner. The scattering influence of relative position change between adjacent units, (c) and (d) SEs comparisons of identical two cylinders and composite units when varying $p$ and $\Delta y$, respectively. The red dashed frame denotes a composite unit. In the insets, the red and green arrows correspond to the positive $x$ and $y$ directions, and black arrows represent the moving directions of adjacent particles. The minimal face spacing in the $x$ (or $y$) direction is $p$ (or $h$), and the center spacing in the y-direction is $\Delta y$.

## 5. Perturbation immunity

Next, we will discuss the EM responses of the bilayer metasurface facing possible disturbers in practical applications. In practice, the stacked metasurfaces will inevitably be affected by interlayer misalignment and position disorder perturbation [42,45,64-67]. These lead to the stacked metasurfaces having to undergo difficulties in high-precision large-area fabrication and applications in large vibration and temperature variation scenarios. Hitherto, the rotation [68] and flip [69] disorder immunities in stacked metasurfaces have been discussed, however, there is still no effective way to solve the problems of interlayer misalignment and positional disorder perturbation.

### 5.1 Layer misalignment

Figure 8(a) shows the R and T curves of the bilayer metasurface corresponding to point B whereas the two MMs are misaligned. The inset illustrates the bottom MM is moving in the x-direction, where the center spacing of two cylinders of the composite unit is $\Delta x$ and the variation range is within one unit cell. Here, the $\eta$ is about 1.11. Thus, with $\Delta x$ and EM dissipation increased, the T transfers from the original maximum value to the null value of point F then to the second local maximum value of point G, while then the R shows the opposite behavior.

Figures 8(b)-8(d) demonstrate the **E** distributions at points E, F, G, respectively, where the **k** is the normally incident wave vector. The numbers around each cylinder in Figs. 8(b) and 8(d) represent six localized regions. Note that, $i$ ($i=1,\cdots,6$) and $i'$ are symmetrical about the horizontal axis when $\Delta x=0$. The R is suppressed and T is enhanced for occurring the coupling between two cylinders of the composite unit in Figs. 8(b) and 8(d), however, the T is suppressed and R is enhanced due to nearly no coupling in Fig. 8(c). Importantly, symmetric region coupling (i.e. 5 and 5' in Fig. 8(b)) has a much stronger ability to suppress R and enhance T than the asymmetric coupling one (i.e. 4 and 5', 5 and 6' in Fig. 8(d)). Therefore, the bilayer metasurface maintains a good transmission as long as the interlayer coupling occurs even if there have layer misalignment.

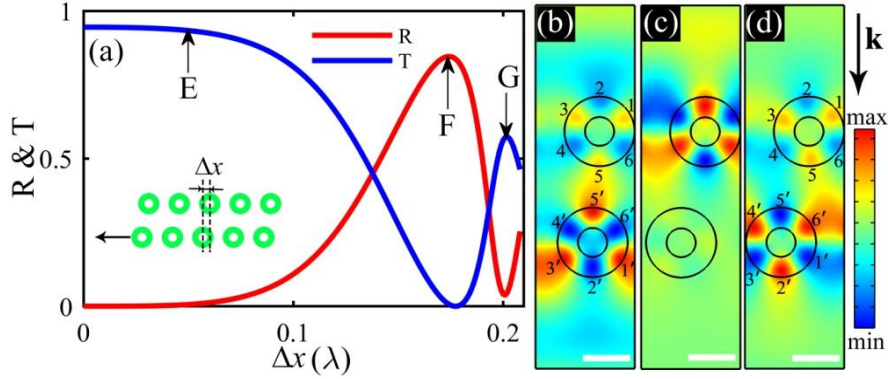

Figure 8. (a) R and T spectra of the bilayer metasurface corresponding to point B with center misalign spacing $\Delta x$ in the *x*-direction. (b)-(d) Electric field **E** distributions corresponding to points E, F, and G, respectively.

**5.2 position disorder perturbation**

Now we introduce disordered position perturbation into finite wide bilayer metasurface and single-layer MM, which consist of 51 composite units and cylinders respectively. The bilayer metasurface corresponds to point B and the plane wave normally incident from above (the directions $x \perp \mathbf{k}$ and $y /\!/ \mathbf{k}$). Under disorder perturbation in the *x*-direction, Fig. 9(a) shows an arbitrary cylinder initially located at the red ring that probably moves to the left or right green ring with the distances of $\Delta x_1$ and $\Delta x_2$. For all cylinders, position disorder degree in the *x*-direction is defined as

$$\sigma_x = \max\{\frac{|\Delta x_1| + |\Delta x_2|}{p_0}\}. \quad (9)$$

Similarly, Fig. 9(b) describes disorder perturbation in the *y*-direction and

$$\sigma_y = \max\{\frac{|\Delta y_1| + |\Delta y_2|}{h_0}\}. \quad (10)$$

Figure 9(c) illustrates the schematic of the bilayer metasurface with the same $\sigma_x$ in both layers, and Figure 9(d) shows the case with position perturbation in the *y*-direction.

Figure 9(e) shows the numerical results of the response factor ($\rho$) of the metasurfaces to different $\sigma_x$. For convenience, the $\rho$ is defined as the major EM response ratio of the disordered to the periodic metasurfaces, i.e. the ratio of R for MM and the ratio of T for bilayer metasurface. Therefore, the greater $\rho$ indicates the smaller change of EM responses of metasurfaces. Random position arrangement only in the bottom layer is RB, only in the top is RT, both in the bottom and top is RBT, and in the single-layer metasurface is RS. Each point and related error bar are the average value and standard error of $\rho$ of 10 samples with the same $\sigma_x$, respectively. Increasing $\sigma_x$, the $\rho$ of RS decreases rapidly, while the bilayer metasurfaces drop more slowly and exhibit much larger $\rho$. Furthermore, the reduced

0th order transmission of RBT mainly transfers into the non-0th order radiations, while the reduced 0th reflection of RS mainly transfers into the 0th order transmission and the non-0th order radiations [see Figs. 10(a) and 10(b)]. In bilayer metasurfaces, the RB and RT are better than the RBT, the RB is consistent with the RT which indicates the disorder in any layer of the bilayer metasurface possesses almost the same EM response. Therefore, the bilayer metasurface has a better performance robust to position perturbation in the $x$-direction.

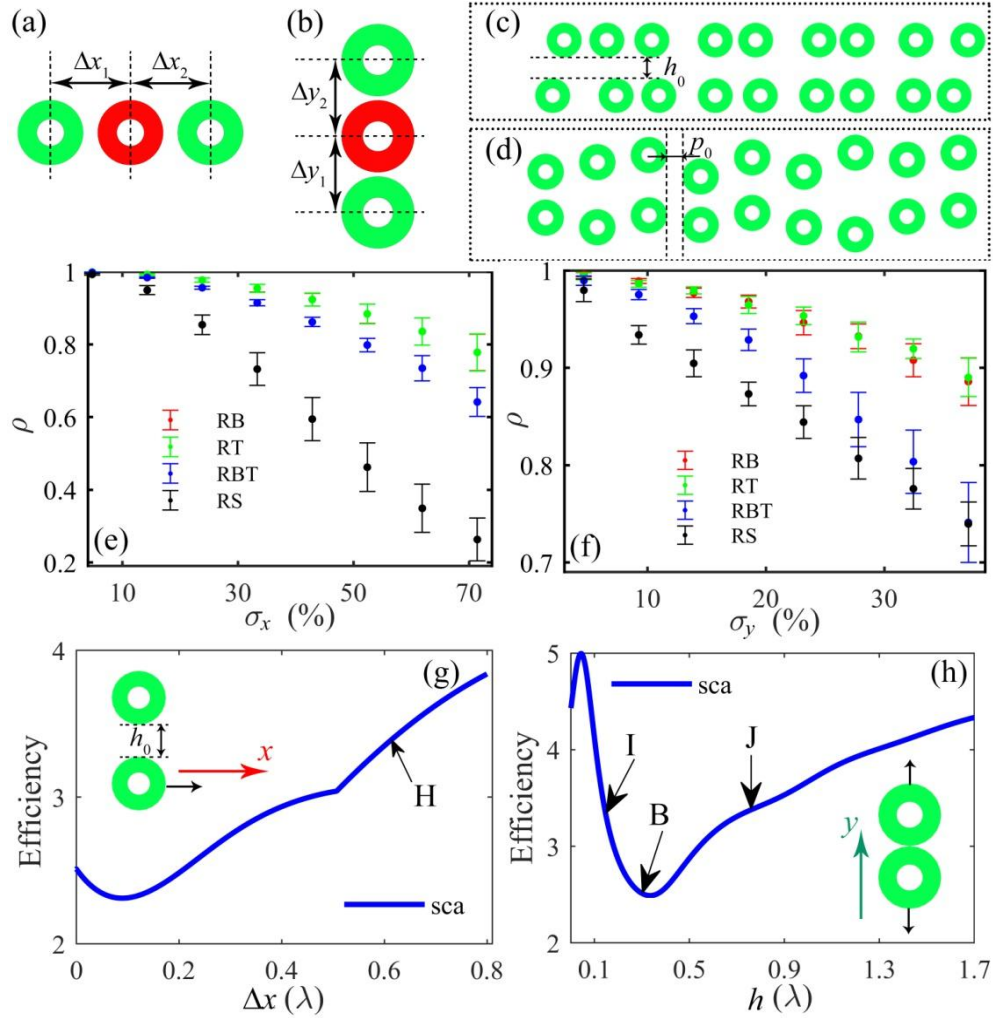

Figure 9. Schematics of the position perturbation (a) in the $x$-direction and (b) in the $y$-direction with the initial positions marked by red rings. The bilayer metasurface at point B with disorder perturbation (c) in the $x$-direction keeping $h_0$ and (d) in the $y$-direction keeping $p_0$. (e) $\rho$ of bilayer and single-layer metasurfaces with different disorder perturbation $\sigma_x$ in the $x$-direction. Each point and the related error bar correspond to the mean and standard deviation of 10 samples with the same $\sigma_x$, respectively. (d) $\rho$ concerning disorder perturbation $\sigma_y$ in the $y$-direction. (g) and (h) SEs of two identical cylinders of a composite unit when $\Delta x$ and $h$ vary.

As a verification, we measure the T of the disordered metasurfaces with $\sigma_x$=71.43% in the experiment. In Fig. 4, the T of the RS is the black circle, the RB is the blue dot, the RT is the green dot, the RBT is the black dot. We approximately estimate the R of RS is 1-T, as the absorption of the cylinders is almost lossless. As a result, the reflection of single-layer MM decreases significantly, but the bilayer metasurfaces can maintain high transmission. The $\rho$ of the disordered metasurfaces are shown in another inset at the bottom of Fig. 4. Therefore, the experimental results agree well with the numerical simulation, i.e. the high transmission of bilayer metasurfaces has better robustness.

Similar to Fig. 9(e), Fig. 9(f) shows the $\rho$ of the metasurfaces to different $\sigma_y$. The best performing RB and RT change uniformly, and the RBT is still better than RS. However, for larger $\sigma_y$, RBT is no longer better than RS due to the larger spacing destroying the interlayer coupling that induces the high transmission. As a result, the RBT will generate the obvious 0th order reflection [see Fig. 10(c)].

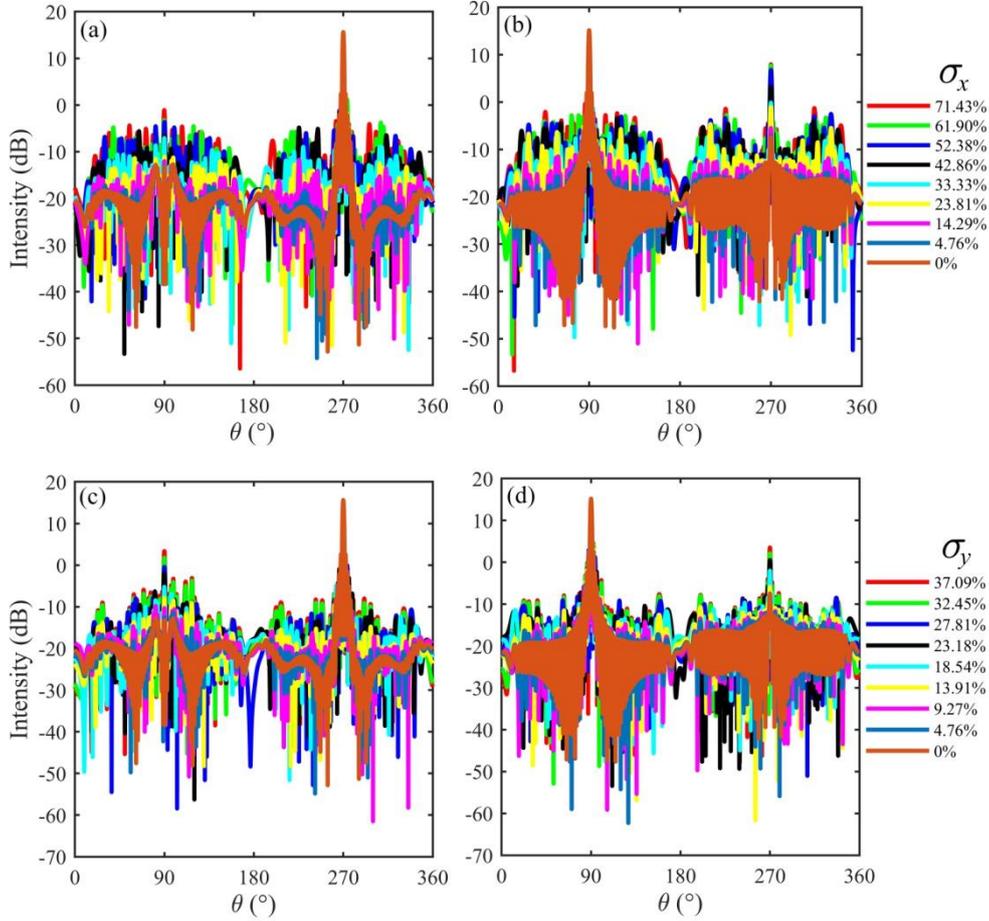

Figure 10. Far-field electric field intensity in all directions of (a) RBT concerning $\sigma_x$, (b) RS concerning $\sigma_x$, (c) RBT concerning $\sigma_y$, and (d) RS concerning $\sigma_y$.

To understand the robustness of the bilayer metasurface, we discuss the scattering of two cylinders of the composite unit with positional perturbations. In Fig. 9(g), the top cylinder is fixed and the bottom cylinder moves along the positive x-direction, where the spacing $h_0$ is fixed. The SE at point H equals that at point A in Fig. 5(a). Note that the same SE curve will be achieved when the bottom moves along the negative x-direction. Figure 9(h) shows the SE of the two cylinders with relative move in the y-direction. The SEs at points I, J are equal to that of point A in Fig. 5(a). Point B is very close to the minimum of SE, which indicates symmetrical coupling at point B can best suppress scattering. Therefore, the bilayer metasurface is more robust than the single-layer case because the composite unit can maintain the relatively lower scattering with a certain degree of position perturbation, especially in the x-direction which has a greater tolerance.

## 6. Conclusions

We have presented a prototype of a tunable bilayer metasurface stacked by two identical MMs consisting of infinitely long air-$Al_2O_3$ concentric cylinders in a parallel arrangement. The MM is a high reflection with large-scale near-field electric enhancement, where each cylinder excites significant ED resonance and other magnetic multipole modes. The bilayer metasurface exhibits dynamical and near-equal intensity interconversion between high reflection and high transmission by flexible adjusting interlayer spacing. Reducing the spacing promotes the interlayer coupling leading to the suppression of the scattering and the achievement of high transmission, however, increasing the spacing weakens the interlayer coupling leading to the enhancement of scattering and the realization of high reflection. Moreover, symmetrical coupling of the two cylinders in the composite unit benefits the transmitted maximization. The transmission of the bilayer metasurface has good robustness due to the composite unit can maintain the scattering suppression whether the disordered position perturbation or the relative movement between the adjacent units. The microwave experimental measurement has verified the numerical result. The next step will be to perform further investigation of the other new features and applications of bilayer metasurfaces stacked by MMs. Our work demonstrates that stacking MMs is a simple and efficient way to design tunable metasurfaces, reveals the scattering properties of the stacking MMs, may shed new light on dynamic components for communication, sensing, logic photonic chips, and could promote us to design robust metasurfaces that can adapt to complex environments.

## Acknowledgments

This work was supported by the National Natural Science Foundation of China (Nos. 11734012, 12074267, 62105213), the Science and Technology Project of Guangdong (No. 2020B010190001), the Guangdong Basic and Applied Fundamental Research Foundation (No. 2020A1515111037), and the Shenzhen Fundamental Research Program (No. 20200814113625003).

## Disclosures

The authors declare no competing financial interest.